\documentclass[a4paper,11pt]{article}
\usepackage{pos}
\usepackage{wrapfig}
\usepackage{lipsum}
\usepackage{xspace}  
\usepackage{notes2bib}
\usepackage{braket}  %
\usepackage{caption}
\usepackage{subcaption}
\DeclareCaptionLabelFormat{withfigure}{Figure~#1 \arabic{figure}#2:}
\let\oldcite\cite
\renewcommand{\cite}[1]{\mbox{\oldcite{#1}}}

\newcommand{\fanto}{{\sf Fant\^omas}\xspace}  %

\newcommand{\xfitter}{{\mbox{\sf xFitter}}\xspace}  %
\newcommand{\FantoPDF}{{\mbox{\sf FantoPDF}}\xspace}  %

\newcommand{\fhead}[1]{\vspace{8pt}\goodbreak \noindent \textbf{#1}:\vspace{5pt}\nobreak\par}

\def\beq{\begin{equation}}
\def\eeq{\end{equation}}
\def\bea{\begin{eqnarray}}
\def\eea{\end{eqnarray}}

\def\smu{\affiliation[a]{Department of Physics, Southern Methodist University,
    Dallas, TX 75275-0175, U.S.A.}}

\def\unam{\affiliation[b]{Instituto de F\'isica,
  Universidad Nacional Aut\'onoma de M\'exico, \\ \qquad 
  Apartado Postal 20-364,
  01000 Ciudad de M\'exico, Mexico}}

\def\msu{\affiliation[c]{Department of Physics,
Michigan State University
East Lansing, MI 48824}}

\newcommand{\orcidFO}{0000-0001-6799-2436} %
\newcommand{\orcidPN}{0000-0003-3732-0860} %
\newcommand{\orcidAC}{0000-0001-8906-2440} %
\newcommand{\orcidMAX}{0009-0003-0139-4072} %
\newcommand{\orcidLK}{0009-0007-5639-0350} %

\newcommand{\orcid}[1]{\,\href{https://orcid.org/#1}{\includegraphics[width=9pt]{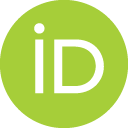}}}

\title{\fanto: An analysis of parton distributions in a pion with Bézier parametrizations}
\ShortTitle{\fanto: An analysis of PDFs in a pion with Bézier parametrizations}

\author[a]{Lucas Kotz\orcid{\orcidLK}} 
\author[b]{Aurore Courtoy\orcid{\orcidAC}} 
\author[c]{Pavel Nadolsky\orcid{\orcidPN}}
\author*[a]{Fredrick Olness\orcid{\orcidFO}}
\emailAdd{olness@smu.edu}
\author[c]{Maximiliano Ponce-Chavez\orcid{\orcidMAX}}

\smu
\unam
\msu

\abstract{
We systematically explore the parametrization dependence of the Parton Distribution Functions (PDFs) to better quantify the true uncertainty from global QCD analyses.
To achieve this, we employ a novel technique that automates the generation of polynomial parametrizations for PDFs using Bézier curves. This technique is implemented in a C++ module, named Fantômas, which is now included within the xFitter program.
As an example, we examine the charged pion PDF. Our analysis reveals that the sea and gluon distributions in the pion are strongly correlated, and PDF solutions featuring a vanishing gluon and a large quark sea are still experimentally allowed.
} %

\FullConference{XXXII International Workshop on Deep Inelastic Scattering and Related Subjects (DIS2025)\\
24-28 March, 2025\\
Cape Town, South Africa\\}

\begin{document}
\maketitle

\def\figPDF{

\begin{wrapfigure}{R}{0.5\textwidth}
\centering
\null\vspace{-50pt}
\includegraphics[width=0.45\textwidth]{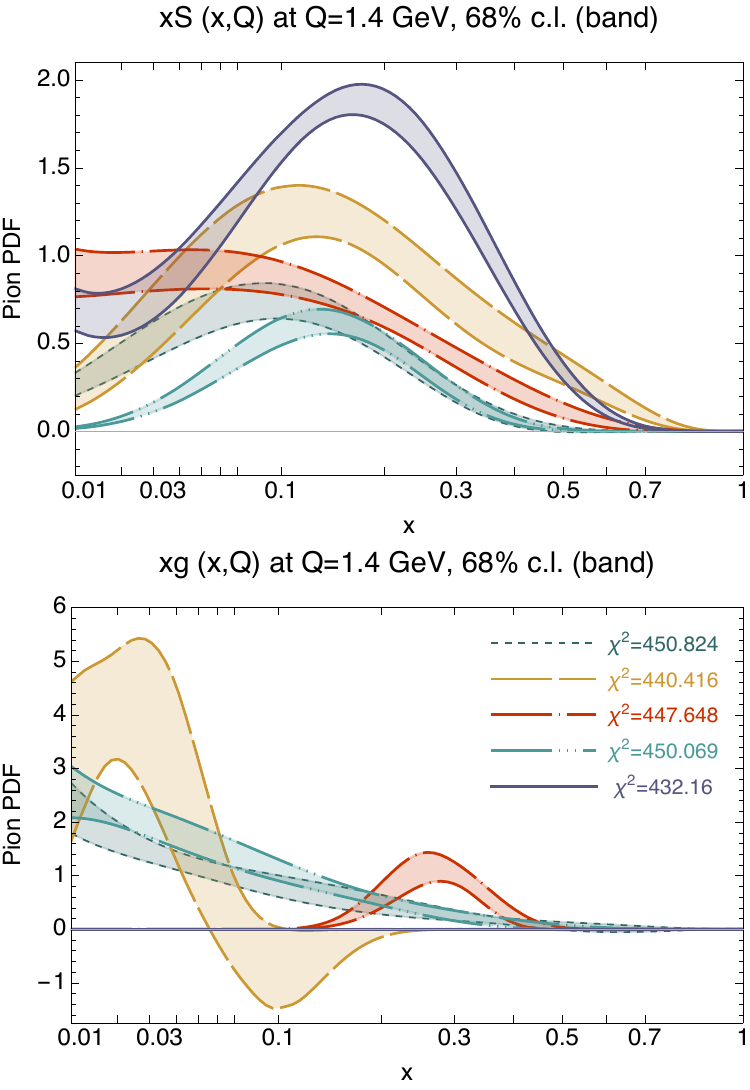}
\null\vspace{-10pt}
\caption{
Final five
sets of the \fanto  PDF combinations
with their respective uncertainty bands.
\\[-5pt] \null \hrulefill
\null\vspace{-20pt}
}  %
\label{fig:one}
\end{wrapfigure}

} %
\def\figCOMBO{
\begin{figure}[t]
     \centering
        \null\vspace{-10pt}
     \begin{subfigure}[t]{0.55\textwidth}
         \centering
        \includegraphics[width=\textwidth]{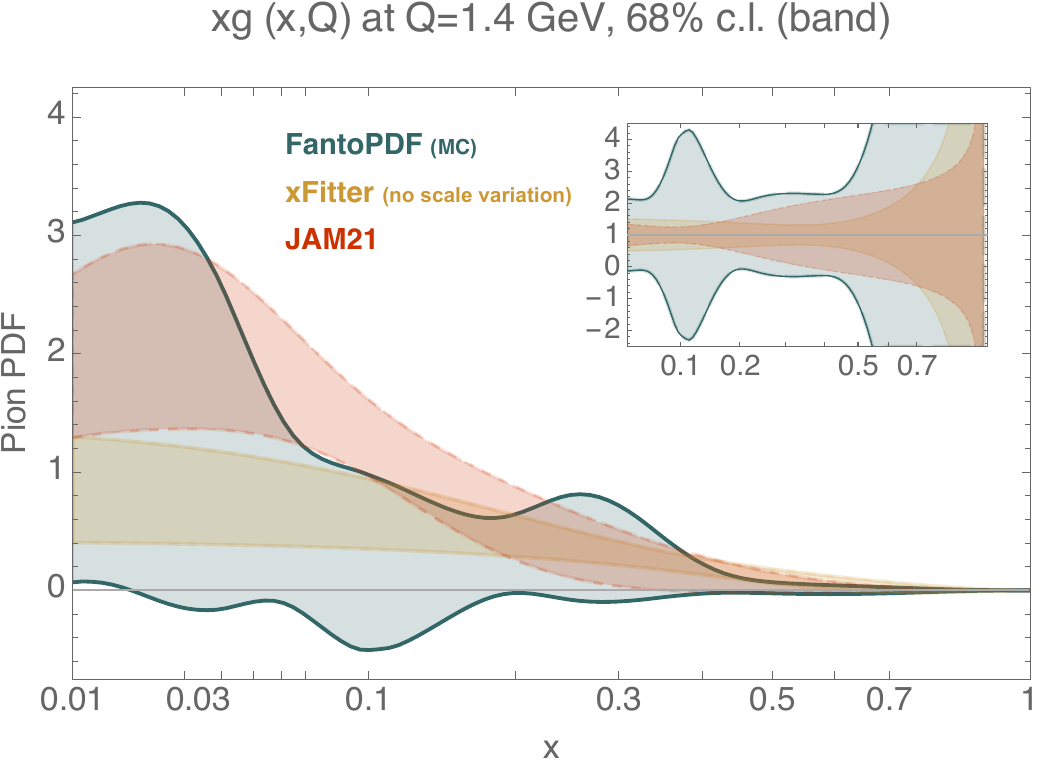}
        \null\vspace{-10pt}
        \caption{ 
        The gluon PDF of the final \fanto ensemble compared to JAM21 and \xfitter results,
          at $Q=1.4$ GeV. For the \FantoPDF set, the
          $68\%$ CL of the MC output is shown.  The inner frame shows the ratio to the central value of
          each set -- symmetric uncertainties are used for all three sets.
          }
         \label{fig:gluon}
     \end{subfigure}
     \hfill
     \begin{subfigure}[t]{0.42\textwidth}
         \centering
         \includegraphics[width=\textwidth]{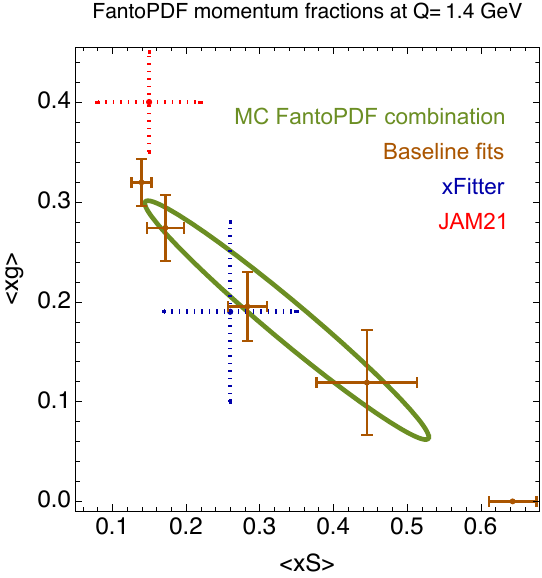}
        \null\vspace{-10pt}
        \caption{
        Momentum fractions for the sea and gluon PDFs obtained in the five baseline fits (orange-brown) and the MC \FantoPDF combination (green ellipse) at $Q_0$. JAM21 and \xfitter results are displayed in red and blue, respectively.}
         \label{fig:mom}
     \end{subfigure}
    \null\vspace{-10pt}

\end{figure}
} %
\null\vspace{-30pt}
\fhead{The \fanto Framework for PDF Analyses}

We present the Fantômas program~\cite{Kotz:2023pbu},
a~C++ package developed to parametrize a variety of Parton Distribution Functions (PDFs) using polynomial shapes realized by Bézier curves.%
This choice is theoretically grounded: Bézier parametrizations can approximate an arbitrary continuous function with desired accuracy, a consequence of the Stone-Weierstrass approximation theorem.

The Fantômas program~\cite{Kotz:2023pbu,Kotz:2025une} 
is incorporated into the public code xFitter~\cite{xFitterwebsite}. We employed this framework to gain valuable insights into the pion Parton Distribution Functions (PDFs).

In contrast to the extensively studied proton PDFs, the pion PDFs are significantly less constrained, and we anticipate they will exhibit substantial parameterization dependence. Thus, the charged pion presents an ideal and timely case for examination using the \fanto framework.
We will use the \xfitter study~\cite{Novikov:2020snp} as our starting point and build upon it to extend the analysis further using the  HERA leading-neutron production in deeply inelastic scattering
(DIS) as studied by the JAM collaboration~\cite{Barry:2018ort,Barry:2021osv}.

\fhead{The \fanto Parametric Form}

We extend the usual parameterization of the hadron PDFs at the initial scale $Q_0$ 
using the following functional form:
\begin{eqnarray}
    x\, f_i(x, Q_0^2)&=& 
    F_i^{car} \times F_i^{mod} 
    =
    \left\{ A_i x^{B_i} (1-x)^{C_i}  \right\}
    \times
    \left\{ 1+ {\cal B}^{(N_m)}[y(x)]  \right\}
\label{eq:fanto_FF_0}
    \label{eq:fanto_FF}
\end{eqnarray}
where,  $i$ runs over all partons. 
$F_i^{car}(x)$ is the   {\it ``carrier function''} which contains 
the normalization coefficients $A_i$, chosen to satisfy the number and momentum sum rules. 
The powers $x^{B_i}$ and $(1-x)^{C_i}$ ensure appropriate limits as $x\to 0$ and $1$.

The {\it ``modulator function''}, $F_i^{mod}(x)$,  provides the additional flexibility to explore a broader range of parameterization forms than the usual PDF analyses. 
In our methodology, the  ${\cal B}$ polynomial is chosen to be a B\'ezier curve. %
Specifically,  ${\cal B}^{(N_m)}[y(x)]$ is B\'ezier   curves  of order $N_m$ 
and the functional argument $y(x)$ allows for a possible rescaling of the $x$-variable. They differ with usual parametrizations  in that their generation can be associated with a set of hyperparameters $(N_m, y(x), \underline{x})$, where $\underline{x}$ represents the abscissa of the control points defining the shape through  shifts of the coordinates $\delta \underline{P}(\underline{x})$ as optimized parameters.
In the limit  ${\cal B}^{(N_m)}=0$, we recover a traditional  baseline parametrization $F_i^{car}(x)$.

\fhead{\fanto Pion Fit}
\nobreak

The \xfitter pion PDF analysis included 
Drell-Yan (DY) pair production by $\pi^-$ scattering on a tungsten target by
 E615 (140 data points)  and NA10 (140 data points), as well as
 prompt-photon ($\gamma$) production in $\pi^-$ and  $\pi^+$ scattering on a tungsten target by WA70 (99 data points).  
 The kinematical coverage of those data is most sensitive to the 
 pion PDFs at large $x$ values. DY
 processes are most sensitive to valence distributions, with the gluon contributing mainly through the DGLAP evolution. The prompt-photon data
 provide additional constraints on the gluon distribution above
 $x\gtrsim 0.1$. 

To better facilitate the separation of the quark sea ($S$) and the gluon ($g$) PDFs at small-$x$ ($x<0.1$), we followed the proposal of the JAM collaboration~\cite{McKenney:2015xis, Barry:2018ort} and included pion Deep Inelastic Scattering (DIS) 29 data  points from H1 at HERA. %
Our choice of data differs largely from JAM’s-- JAM's fit use as much LN data as DY, while Fant\^omas's has 10 times more DY data.

\fhead{The PDF Fits}
We proceed with an in-depth investigation of the functional form
dependencies enabled by the Fantômas methodology.
We determine the PDFs at the next-to-leading order (NLO) in the QCD coupling strength $\alpha_s$, and assume the same flavor composition of pion PDFs at the initial scale $Q_0^2 =1.9\mbox{ GeV}^2$ as in the \xfitter study~\cite{Novikov:2020snp}. In addition to the gluon PDF $g(x)$, we introduce the total valence and sea quark distributions, $V(x)$ and $S(x)$, and determine the PDFs for individual flavors on the assumptions that the PDFs for the two constituent (anti)quarks are the same in $\pi^+$ and $\pi^-$, and the light quark sea is flavor-blind.

In the fits performed on the complete data set, we independently vary the
Bézier curve's hyperparameters, including the order, which lies between 0 and 3;   altogether, the fits presented have a minimum of 5
and a maximum of 13 free parameters.

Within the \fanto framework we generate ${\sim}100$ baseline PDF sets 
as outlined in Ref.~\cite{Kotz:2023pbu}.
The expected variance of the $\chi^2$ for $N_{\rm pts}=408$ and
varying $N_{\rm par}$ between 7 and 13 amounts to $\delta
\chi^2 \simeq 30$ at the $1\sigma$ level~\cite{Kovarik:2019xvh}; most of the PDF sets are well within this range. 

The spread of these baseline PDFs quantifies the parametrization dependence that contributes to the epistemic uncertainty~\cite{Courtoy:2022ocu}.
We want to combine the epistemic uncertainty (characterized by the parameterization dependence)
with the aleatoric uncertainty (characterized by stochastic fluctuations of data) to obtain 
a total PDF uncertainty. 

\fhead{The PDF Uncertainties}
In many practical applications with limited parameter dependence, a useful estimate of epistemic uncertainty can often be obtained by sufficiently sampling acceptable PDF models \cite{Courtoy:2022ocu}.
This procedure is more direct than introducing either the global or dynamic tolerance.

\noindent
We follow this paradigm in three steps to estimate the total PDF uncertainty.

\textbf{1) Epistemic Uncertainty:} 
We select five out of ${\sim}100$ explored baseline PDF
configurations that yield a good $\chi^2$ and whose central fits
collectively cover a broad span of possible PDF magnitudes 
to estimate the epistemic uncertainty.

\textbf{2) Aleatory Uncertainty:} 
For each of these five selected models, we
generate an uncertainty band using the Hessian method with $\Delta \chi^2=1$ to estimate their aleatory uncertainties. %
These five PDFs with error bands\footnote{%
The combined ensemble is generated using the {\sf mp4lhc} package~\cite {Gao:2013bia}.
For each of the five input Hessian ensembles shown in
Fig.~\ref{fig:one}, which contain $2N_{\rm eig}$
eigenvector PDFs, we produce 100 Monte Carlo replicas.  These replicas
are generated using a linear sampling procedure based on symmetric
eigenvector variations.
} %
are displayed in Fig.~\ref{fig:one}.

\textbf{3) Combined Uncertainty:} 
Finally, we generate a combined PDF error ensemble
from these five bands using the METAPDF method to achieve a more
robust total uncertainty estimate~\cite{Gao:2013bia}. The METAPDF package was updated for arbitrary hadrons in Ref.~\cite{Kotz:2025lio}.
The resulting ensemble of pion error PDFs is named ``\FantoPDF''
and the result for the gluon is displayed in Fig.~\ref{fig:gluon}.

To summarize, the five fits that form the final \FantoPDF ensemble
were selected for their distinct shapes, ensuring a broad coverage of
the parameter space explored across the ${\sim}100$ baseline configurations
tested. These selected fits
are then combined using the {\sf METAPDF} technique, which effectively
captures the dominant parametrization uncertainty, even though it
excludes a few extreme baseline solutions exhibiting negative PDFs.

\fhead{Observations}
\figPDF

Upon examination of Figures~\ref{fig:one} and~\ref{fig:gluon},  the picture that emerges is that the flexibility of the functional form strongly affects conclusions about the experimentally allowed momentum fractions, increasing their ranges compared to fits with fixed parametrization forms.
Specifically, the spread of the solutions in Fig.~\ref{fig:one} quantifies the
parametrization dependence that contributes to the epistemic uncertainty.
We observe that the sample five PDF sets span a much larger range collectively than any individual set; this is the case even if we were to increase the  $\Delta \chi^2$ with a tolerance factor.  

In Fig.~\ref{fig:gluon}, we present our \FantoPDF for the gluon, and compare with the results from xFitter and JAM. 
The inner frames of Fig.~\ref{fig:gluon} show the ratios to the central PDF of the corresponding sets.
We observe that the \fanto gluon uncertainty band is significantly enlarged across the entire kinematic range of $x$ compared with the other sets.
In particular, we note that, due to this expanded uncertainty band, a scenario with a small gluon PDF is now fully consistent and contained within the allowed uncertainty range. 
The comparative difference in the mid-$x$ range might be sensitive to the choice of the model for the pion flux in the description of the Leading Neutron (LN) data. Furthermore, it corresponds to the transition from the LN and the pion-induced Drell-Yan (DY) data set, while still being bridged by the prompt-photon data. 
The \fanto fit is the first to include all three processes, on
top of the \fanto framework that accounts for uncertainty sources
beyond the aleatory one.

\fhead{Gluon and Sea PDF Correlations}
\figCOMBO

The sea and gluon PDFs are intrinsically connected through the DGLAP evolution equations. This coupling allows variations in one flavor (e.g., the gluon) to be compensated by changes in the other (the sea quarks). Their interplay is further constrained by the momentum sum rule.

To highlight this relationship, we can compute the integrated momentum fractions $\langle xf\rangle$ for the gluon and sea distributions.
The valence sum rule governs the momentum fraction for $\langle x V\rangle$, while the momentum sum rule relates the sea and gluon momentum fractions. 
In Fig.~\ref{fig:mom} 
we plot the momentum fractions for the five baseline fits as well as for the final \FantoPDF combination. We also superimpose the respective momentum fractions from the JAM'21 \cite{Barry:2021osv} and \xfitter \cite{Novikov:2020snp} analyses.
The long $\braket{xS}$-$\braket{xg}$ correlation ellipse (green) for the final \fanto combination in Fig.~\ref{fig:mom} demonstrates the strong correlation between the gluon and sea distribution.
Notably, the scenario in which the gluon carries above 40\% of the momentum is disfavored in the \fanto analysis, as it results in an overly low (anti)quark sea that undershoots the 
NA10 and WA70
cross sections. 
Conversely, solutions with a low or even zero gluon PDF often improve the description of Drell-Yan data, yielding low $\chi^2$ values. This particular conclusion may be sensitive to both the order of the perturbative QCD calculation,
factor which will require further exploration. On the CTEQ-TEA github
\footnote{\href{https://cteq-tea.gitlab.io/project/00pdfs/\#mesonPDFs}{https://cteq-tea.gitlab.io/project/00pdfs/\#mesonPDFs}} 
as well as on the LHAPDF page, we provide the Fanto10\_n15 PDF set. It is a 15-member eigenvector set that also includes uncertainties from nCTEQ15 nuclear PDFs entering in the fit to the DY data~\cite{Kotz:2025lio}.

\fhead{Conclusion}

This study builds upon the insightful NLO phenomenological analysis of charged pion PDFs previously performed by JAM~\cite{Barry:2018ort,Barry:2021osv} and xFitter~\cite{Novikov:2020snp} collaborations. 
We employ the \fanto framework to systematically incorporate both epistemic (knowledge-based) and aleatory (random-variability) uncertainties, enabling the construction of a robust \FantoPDF set. The increased flexibility of the functional form significantly affects the inferred uncertainty range of the resulting PDFs and the allowed momentum fractions. 
This enhanced flexibility leads to broader, and likely more realistic, uncertainty bands compared to fits constrained by fixed parameterization forms.

Accurate quantification of the total uncertainty is essential,  as the kinematic range covered by experimental data, order of the calculation (still NLO in all analyses), and modeling of  the LN pion flux and nuclear uncertainties directly 
affects comparisons with predictions from nonperturbative QCD calculations. Providing a reliable uncertainty estimate is particularly important now, given the rapid progress in theoretical simulations and the upcoming experimental measurements of both meson and baryon structure.
Raising the accuracy of these components to the next level will help solve the existing puzzle with the pion structure, namely the opposing preferences of the lattice QCD predictions~\cite{Barry:2025wjx} and DY data for a large and small gluon content of the pion, respectively.

\fhead{Acknowledgment}
This work was performed in part at the Aspen Center for Physics, which is supported by National Science Foundation grant PHY-2210452.
This work was also supported by  %
a National Science Foundation AccelNet project, 
by the U.S.\ DoE  Grant No.~DE-SC0010129,
and 
by the  U.S.\ DoE, Office of Science, Office of Nuclear Physics, 
within the framework of the Saturated Glue (SURGE) Topical Theory Collaboration.
PMN is grateful for support from the Wu-Ki Tung Endowed Chair in particle physics.
AC and MPC were further supported by the UNAM Grant No. DGAPA-PAPIIT IN102225.

\bibliographystyle{apsrev4-1}
\bibliography{extra,main} 

\end{document}